\documentclass[%
superscriptaddress,
reprint,
amsmath,amssymb,
aps,
prx,
10pt,
twocolumn,
floats,
]{revtex4-2}

\usepackage[english]{babel}
\usepackage[utf8]{inputenc}
\usepackage[T1]{fontenc}
\usepackage{graphicx}
\usepackage[caption=false]{subfig}
\usepackage{float}
\usepackage{dcolumn}
\usepackage{blindtext}
\usepackage{upgreek}
\usepackage{siunitx}
\usepackage{xcolor}

\DeclareSIUnit\gauss{G}
\usepackage{braket}

\newcommand{\Sup}{\uparrow}
\newcommand{\Sdown}{\downarrow}
\newcommand{\Sgo}{0}
\newcommand{\Sgt}{1}
\newcommand{\SR}{{R_a}}
\newcommand{\SRp}{{R_a'}}
\newcommand{\Saux}{R_c}

\newcommand{\kup}{\ket{\Sup}}
\newcommand{\kdown}{\ket{\Sdown}}

\newcommand{\kgo}{\ket{\Sgo}}
\newcommand{\kgt}{\ket{\Sgt}}
\newcommand{\kR}{\ket{\SR}}
\newcommand{\kRp}{\ket{\SRp}}
\newcommand{\kaux}{\ket{\Saux}}

\newcommand{\prob}{\mathcal{P}}
\newcommand{\probc}{\mathcal{P}_L}

\newcommand{\vv}[1]{\boldsymbol{#1}}

\newcommand{\LA}{_\mathrm{LA}}
\newcommand{\ALA}{_\mathrm{ALA}}

\usepackage{wasysym}
\usepackage{enumitem}
\usepackage{bm}

\begin{document}

\title{Non-destructive optical read-out and manipulation of circular Rydberg atoms}
\author{Y.~Machu}
\author{A.~Durán-Hernández}
\author{G.~Creutzer}
\author{A.~A.~Young}
\author{J.~M.~Raimond}
\author{M.~Brune} 
\affiliation{Laboratoire Kastler Brossel, Coll\`ege de France, CNRS, ENS-Universit\'e PSL, Sorbonne Universit\'e, 11 place Marcelin Berthelot, F-75231 Paris, France}
\author{C.~Sayrin}
\email[Corresponding author: ]{clement.sayrin@lkb.ens.fr}
\affiliation{Laboratoire Kastler Brossel, Coll\`ege de France, CNRS, ENS-Universit\'e PSL, Sorbonne Universit\'e, 11 place Marcelin Berthelot, F-75231 Paris, France}
\affiliation{Institut Universitaire de France, 1 rue Descartes, 75231 Paris Cedex 05, France}
\date{\today}

    \begin{abstract}
Among the thriving quantum computation and quantum simulation platforms based on arrays of Rydberg atoms, those using circular Rydberg atoms are particularly promising. These atoms uniquely combine the strong dipole-dipole interactions typical of Rydberg states with long lifetimes. However, low-angular-momentum ($\ell$) laser-accessible Rydberg levels have been so far mostly used, because circular Rydberg atoms have no optical transitions, hindering their individual detection and manipulation. %
We remove this limitation with a hybrid platform, combining an array of logical laser-trapped circular Rydberg atoms of rubidium with an auxiliary array of Rb ancilla atoms transiently excited to a low-$\ell$ Rydberg level. 
We perform a quantum non-demolition detection of the logical qubit with the ancilla, through the blockade of the ancilla optical excitation induced by a Förster resonance. %
Conversely, we locally manipulate the logical qubit through the excitation of the ancilla. This dual-Rydberg platform is highly promising for quantum computation and simulation. It adds to the circular-atom toolbox the mid-circuit measurements, essential for error correction. More strikingly, it gives access to time correlations in long-term quantum simulations, uniquely accessible to circular Rydberg atoms. 
    \end{abstract}

\maketitle
Rydberg atoms, i.e., atoms with one valence electron excited to a high-principal-quantum-number level, have recently proven to be ideal for quantum computation and quantum simulation~\cite{Browaeys2020, Morgado2021}. The devices typically operate on arbitrary defect-free arrays of individual ground-state atoms laser-trapped in optical tweezers~\cite{Barredo2016, Endres2016}. The atoms are transiently excited to Rydberg levels. Their strong mutual dipole-dipole interaction enables two- or multi-qubit gates~\cite{Graham2022, Bluvstein2022} and quantum simulation of interacting-spin Hamiltonians~\cite{Scholl2021, Ebadi2021}. Remarkable recent achievements include the demonstration of high-fidelity gates~\cite{Ma2023, Scholl2023a, Evered2023}, the operation of a quantum processor~\cite{Bluvstein2024} or the quantum simulation of dynamics of interacting spins~\cite{Chen2025a, Manovitz2025, Gonzalez-Cuadra2025}, with performances ultimately limited by the $\approx \SI{100}{\micro\second}$ lifetime of the laser-accessible Rydberg levels used in these devices. The final read-out of the atomic state relies on fluorescence imaging of a ground state. While it can reach high fidelities and can be made fast~\cite{Bergschneider2018}, it is necessarily destructive for a qubit encoded in Rydberg levels.% 

Hybrid platforms, where ancillae are used to manipulate and detect the logical qubits, are promising to solve this problem. Dual-species experiments propose to encode logical qubits and ancillae in different atomic or molecular species~\cite{Beterov2015, Kaufman2021, Zhang2022, Wang2022}, while coupling them strongly via the excitation to Rydberg levels experiencing a Förster resonance~\cite{Safinya1981, Ryabtsev2010}. This enables, in particular, final or mid-circuit non-destructive quantum state measurements of the logical qubit~\cite{Beterov2015} and multi-qubit gates~\cite{Petrosyan2024}. Recent experimental demonstrations of this architecture employing alkali atoms~\cite{Singh2022, Anand2024} have, in particular, enabled mid-circuit correction of errors~\cite{Singh2023}. These methods require simultaneous advanced manipulation of two different species, increasing significantly the experimental complexity. 

\begin{figure*}[t]
\centering
\includegraphics[width=1.8\columnwidth]{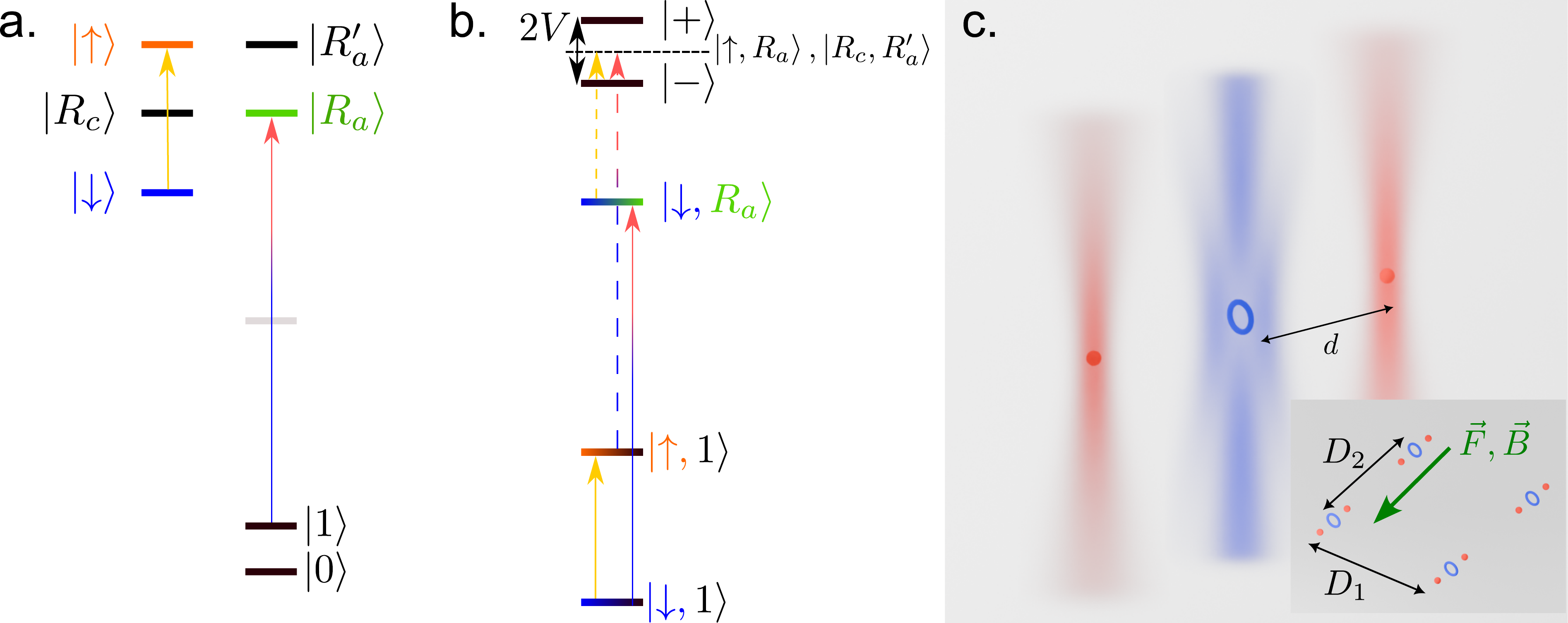}

\caption{\textbf{Level structure and experimental geometry.} \textbf{a.} Simplified structure of non-interacting Rb logical and ancilla atomic levels, not to scale. The logical atom (left) is initially excited to the circular Rydberg level with principal quantum number $n=52$, $\kdown$. A two-photon MW transition (yellow arrow) couples it to $\kup$ ($n=54$ circular state). The ancilla (right) is excited from its ground state $\kgt\equiv\ket{5S_{1/2}, F=2, m_F=2}$ to the Rydberg level $\kR\equiv\ket{45S_{1/2}, m_J = 1/2}$ by a two-photon laser excitation (blue-red-gradient arrows) through the $6P_{3/2}$ intermediate level (gray line). The state $\kgo$ is any ground state in the $\ket{5S_{1/2}, F=1}$ manifold. At the Förster resonance, the energy difference between $\kRp\equiv\ket{45P_{3/2}, m_J=3/2}$ and $\kR$ matches that between $\kaux$ and $\kup$. \textbf{b.} Simplified level structure of a pair of interacting logical and ancilla atoms. Both optical and microwave excitations are allowed (solid arrows) from the pair state $\ket{\Sdown,\Sgt}$ but blocked (dashed arrows) from $\ket{\Sup, \Sgt}$ and $\ket{\Sdown,\SR}$ because of the dipole-dipole interaction $V$ that lifts the degeneracy between the $\ket{\pm}$ states. The dashed line indicates the energy of the non-interacting pair states $\ket{\Sup, \SR}$ and $\ket{\Saux, \SRp}$.
\textbf{c.} An ensemble of three atoms, with a logical atom excited to the circular Rydberg level $\kdown$ (blue torus) and two ancillae in the ground state $\kgt$ (red spheres). The ancillae are trapped in Gaussian optical tweezers (red beams) and the logical atom in an optical bottle beam (blue beam). The programmed interatomic distance is $d=\SI{6}{\micro\meter}$. (Inset) We prepare four ensembles distant by ${D_1 = \SI{50}{\micro\meter}}$ and ${D_2 = \SI{60}{\micro\meter}}$. Within an ensemble, the atoms are aligned along the quantization axis, fixed by the parallel electric, $\vv{F}$, and magnetic, $\vv{B}$, fields (green arrow).}
\label{fig:setup}
\end{figure*} 

In this work, we employ different Rydberg levels of the same atomic species, namely Rubidium 87 (Rb). We encode the logical qubit on two long-lived circular Rydberg levels~\cite{Cantat-Moltrecht2020}, while we engineer the ancilla qubit with two hyperfine sublevels of the ground state.  We perform an \emph{ancilla-based measurement} (ABM) of the state of the logical atom using the dipole-dipole interaction with the ancilla, optically excited to a low-angular-momentum ($\ell$) Rydberg level. It possesses a Stark-tuned Förster resonance with one of the circular Rydberg levels. Because the circular states are impervious to optical light but highly sensitive to microwave fields, the two qubits can be separately manipulated with global addressing fields, hence acting as atoms from different species. We demonstrate that the ABM is a quantum non-demolition (QND) measurement~\cite{Thorne1978, Unruh1978}: It projects the state of the logical atom onto the measured state, according to Born probabilities, and leaves an atom initially in an eigenstate of the measurement unaffected. In addition, we use the interaction between the ancilla and logical atoms for local manipulation of the state of the latter. We induce phase flips and spin flips of the logical qubit through local optical control of the ancilla atom only. 

We encode the states of the logical qubits onto the circular Rydberg levels with principal quantum numbers $n=52$ and $n=54$, denoted $\kdown$ and $\kup$, respectively (see Fig~\ref{fig:setup}.a). The ancilla qubit states, denoted $\kgt$ and $\kgo$, correspond to the $\ket{5S_{1/2},F=2, m_F=2}$ ground state and to any state of the $F=1$ hyperfine ground-state manifold, respectively. We use a microwave (MW) field to induce transitions between $\kdown$ and $\kup$ and laser beams to couple $\kgt$ to the Rydberg level $\kR\equiv\ket{45S_{1/2},m_J=1/2}$. The latter resonantly interacts with $\kup$ only, through a Förster resonance. In an electric field of $F=\SI{1.6}{\volt\per\centi\meter}$, the atom-pair state $\ket{\Sup,\SR}$ is degenerate with $\ket{\Saux,\SRp}$, where $\kRp = \ket{45P_{3/2}, m_J=3/2}$ and $\kaux$ is the $n=53$ circular state~\cite{Suppl}. The electric dipole-dipole interaction couples the states $\ket{\Sup,\SR}$ and $\ket{\Saux,\SRp}$ and lifts their degeneracy: The two new eigenstates $\ket{\pm} = (\ket{\Sup,\SR} \pm \ket{\Saux,\SRp})/\sqrt{2}$ are displaced by $\pm V$ (see Fig.~\ref{fig:setup}.b), where the dipole-dipole interaction strength, $V$, scales as $1/d^3$. At first order, the state $\ket{\Sdown,\SR}$ is unaffected by the dipole-dipole interaction. The strong first-order dipole-dipole interaction between $\kup$ and $\kR$ results in the Rydberg blockade~\cite{Lukin2001a} of the optical excitation from $\kgt$ to $\kR$ of an ancilla atom when a nearby logical atom is prepared in $\kup$ (Fig.~\ref{fig:setup}.b). This excitation is possible when the logical qubit is in $\kdown$. Conversely, an ancilla in $\kR$ prevents the microwave transfer between $\kdown$ and $\kup$ for the logical atom. We use these two processes to demonstrate QND measurement and local manipulation, respectively, of the state of the logical qubit.

The experimental setup is described in details in~\cite{Suppl, Ravon2023}. We initially prepare a defect-free array of Rb atoms, laser trapped in focused Gaussian optical tweezers. We arrange the atoms in ensembles of three trapping sites, separated by a distance set to be $d=\SI{6}{\micro\meter}$, and aligned with the quantization axis (see Figure \ref{fig:setup}.c). This axis is fixed by a constant magnetic field $\vv{B}$ and a parallel initially-vanishing electric field $\vv{F}$. The ensembles are far enough from each other to neglect interactions between atoms of different ensembles. 

In every ensemble, we use an atom in the central site (Fig.~\ref{fig:setup}.c) as a logical atom while atoms in the two outer sites act as ancillae. We can choose to operate with one or two ancillae, depending on the experiment we perform. We denote $\ket{\psi_L, \psi_A}\LA$ ($\ket{\psi_A',\psi_L, \psi_A}\ALA$) the state of an atom pair (triplet) with the logical atom and ancilla(e) in $\ket{\psi_L}$ and $\ket{\psi_A}$ or $\ket{\psi_A'}$, respectively. All the atoms are initially prepared in $\kgt$ and held in Gaussian optical tweezers.

For the time being, we consider for simplicity the situation with only one ancilla. The preparation of $\ket{\Sdown, \Sgt}\LA$ begins with the laser excitation of the logical atom to the low-$\ell$ Rydberg level $\ket{52D_{5/2}, m_J=5/2}$. During this step, we turn off the trapping light for the central atom only, so that the ancilla remains in $\kgt$ because of the trap-induced light shifts. Then, the logical atom is transferred to $\kdown$ using a circularization method described in~\cite{Cantat-Moltrecht2020, Ravon2023}. This process leaves $\kgt$ unaltered. A final optional MW $\pi$ pulse on the two-photon $\kdown\to\kup$ transition prepares $\ket{\Sup, \Sgt}\LA$. The logical atoms, in circular Rydberg states, are trapped in optical bottle beams (BoBs)~\cite{Isenhower2009} (Fig.~\ref{fig:setup}.c) that we turn on right after their initial optical excitation~\cite{Ravon2023}, while the ancillae remain trapped in the Gaussian optical tweezers.

At the end of the experimental sequence, we optically detect the state of the circular Rydberg atoms with a destructive, non-repeatable measurement (DM) demonstrated in Ref.~\cite{Ravon2023}. We revert the excitation process to bring the logical atoms in $\kdown$ back to $\kgt$ and recapture them in Gaussian optical tweezers. All spuriously-populated levels and $\kup$ end up in Rydberg states and are expelled by the optical tweezer traps. A final $\SI{25}{\ms}$-long fluorescence imaging detects laser-trapped ground-state atoms, measuring the population in $\kdown$ prior to the DM. We can measure instead the population in $\kup$ by applying a MW $\pi$ pulse on the $\kdown\to\kup$ transition before this measurement sequence. In both cases, we measure the probability to recapture the logical atom in the optical tweezers, and denote it $\probc^{(\Sup)}$ or $\probc^{(\Sdown)}$, the exponent indicating the detected state. %

\begin{figure}
\centering
\includegraphics[width=0.9\columnwidth]{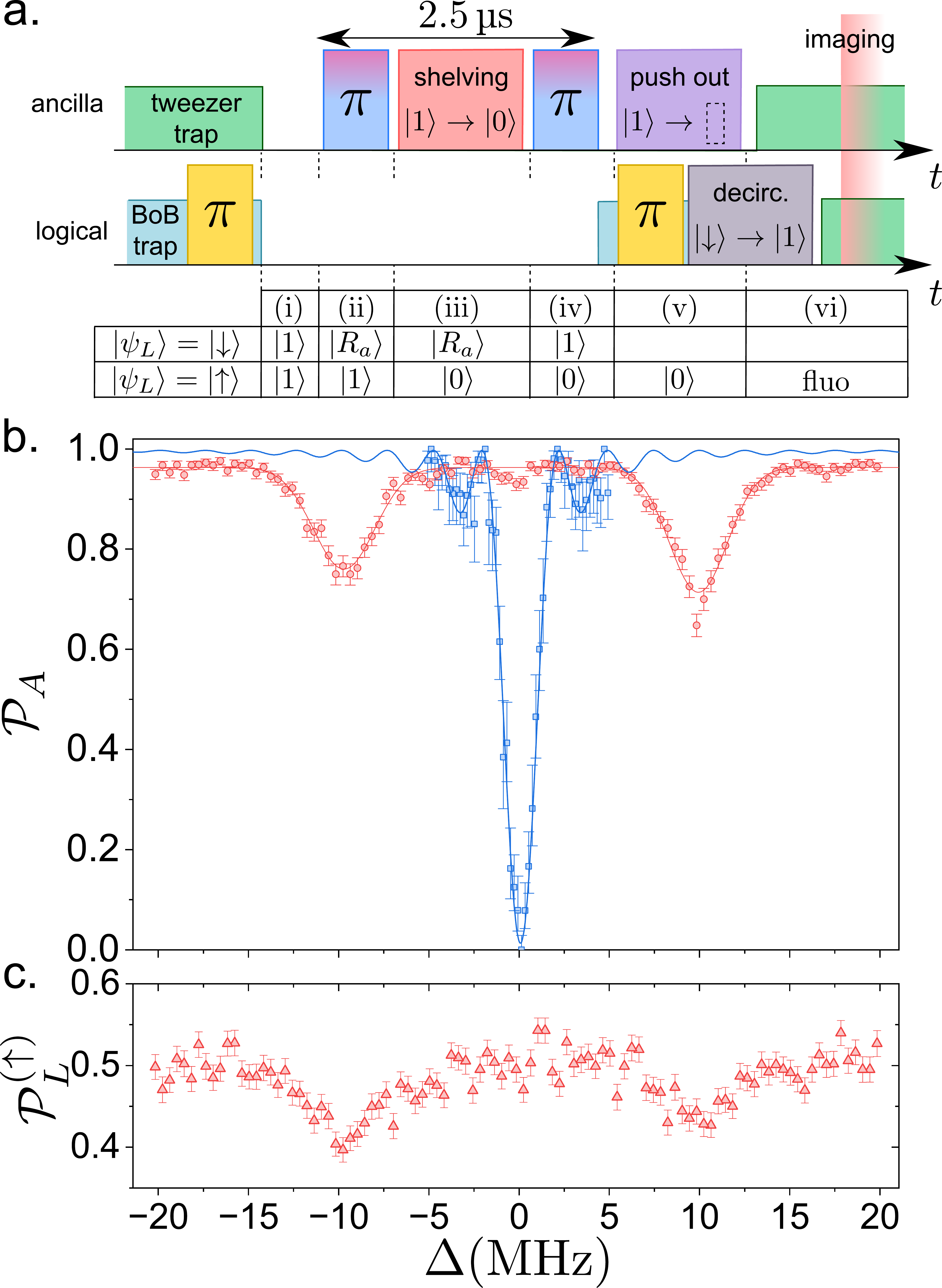}
\caption{\textbf{Ancilla optical excitation spectra.} \textbf{a.} The top (bottom) line is a graphical representation (not to scale) of the sequence of optical (MW) pulses that address the ancilla (logical) atom during the ABM sequence. The boxes filled with a blue-red gradient indicate the Rydberg excitation pulses, the green boxes indicate the time during which we trap the ancilla in the optical tweezers. The bottom line displays the time during which the BoB traps are on (cyan box), the optional MW $\pi$ pulses that transfer $\kdown$ to $\kup$ (yellow boxes) and the decircularization process that brings back the circular state $\kdown$ to $\kgt$ (grey box). The table gives the value of $\ket{\psi_A}$ at every step of the sequence in the cases where the MW pulses on the logical atom is applied (bottom line) or not (top). The dashed lines indicate the steps of the sequence on the graphical representations. \textbf{b.} Excitation spectrum of $\kR$. We plot as a function of the detuning $\Delta$ of the Rydberg excitation lasers from the $\kgt\to\kR$ resonance the probabilities $\prob_{A|\Sdown}$ (blue) and $\prob_{A|\Sup}$ (red) to recapture the ancilla when the logical atom has been prepared in $\kdown$ and $\kup$, respectively. The solid lines are fits to the data of a single coherent excitation profile (blue) or a sum of two Gaussian peaks (red)~\cite{Suppl}. \textbf{c.} Probability to recapture the logical atom when prepared in $\kup$. In panels \textbf{b} and \textbf{c}, the red (blue) points are the result of an average over 400 (50) repetitions, the error bars are statistical (Clopper-Pearson interval). }
\label{fig:Spectrum}
\end{figure}

The \emph{ABM sequence} is sketched in Fig.~\ref{fig:Spectrum}.a: (i) We turn off the traps and (ii) shine the Rydberg excitation lasers on resonance with the $\kgt\to\kR$ transition for $\SI{0.4}{\micro\second}$. If the logical atom is in $\ket{\psi_L} = \kdown$, the optical $\pi$ pulse brings $\ket{\Sdown, \Sgt}\LA$ to $\ket{\Sdown, \SR}\LA$. If $\ket{\psi_L}=\kup$, however, Rydberg blockade prevents the excitation (see Fig.\ref{fig:setup}.b), leaving $\ket{\Sup, \Sgt}\LA$ unaffected. At this stage, the state of the logical atom, $\kdown$ or $\kup$, is mapped onto the state of the ancilla, $\kR$ or $\kgt$. %
(iii) We shelve the atoms left in $\kgt$ into $\kgo$ by optical pumping and (iv) we transfer the atoms excited to $\kR$ back to $\kgt$ with the same optical $\pi$ pulse as in step (ii), avoiding limitations due to the ${\SI{51}{\micro\second}}$ finite lifetime of $\kR$. (v) We expel these atoms in $\kgt$ from the trapping region with a push-out optical beam, (vi) we turn back on the Gaussian optical tweezers to recapture the atoms left in $\kgo$, and we eventually detect all ground-state atoms by a final fluorescence image. The last two steps of the protocol perform a state-selective destructive measurement of the ancilla state $\kgo$. Note that we turn off the BoB traps during steps (ii-iv) to prevent residual light shifts they could induce on the nearby ancilla. %

\begin{figure*}
\centering
\includegraphics[width=1.9\columnwidth]{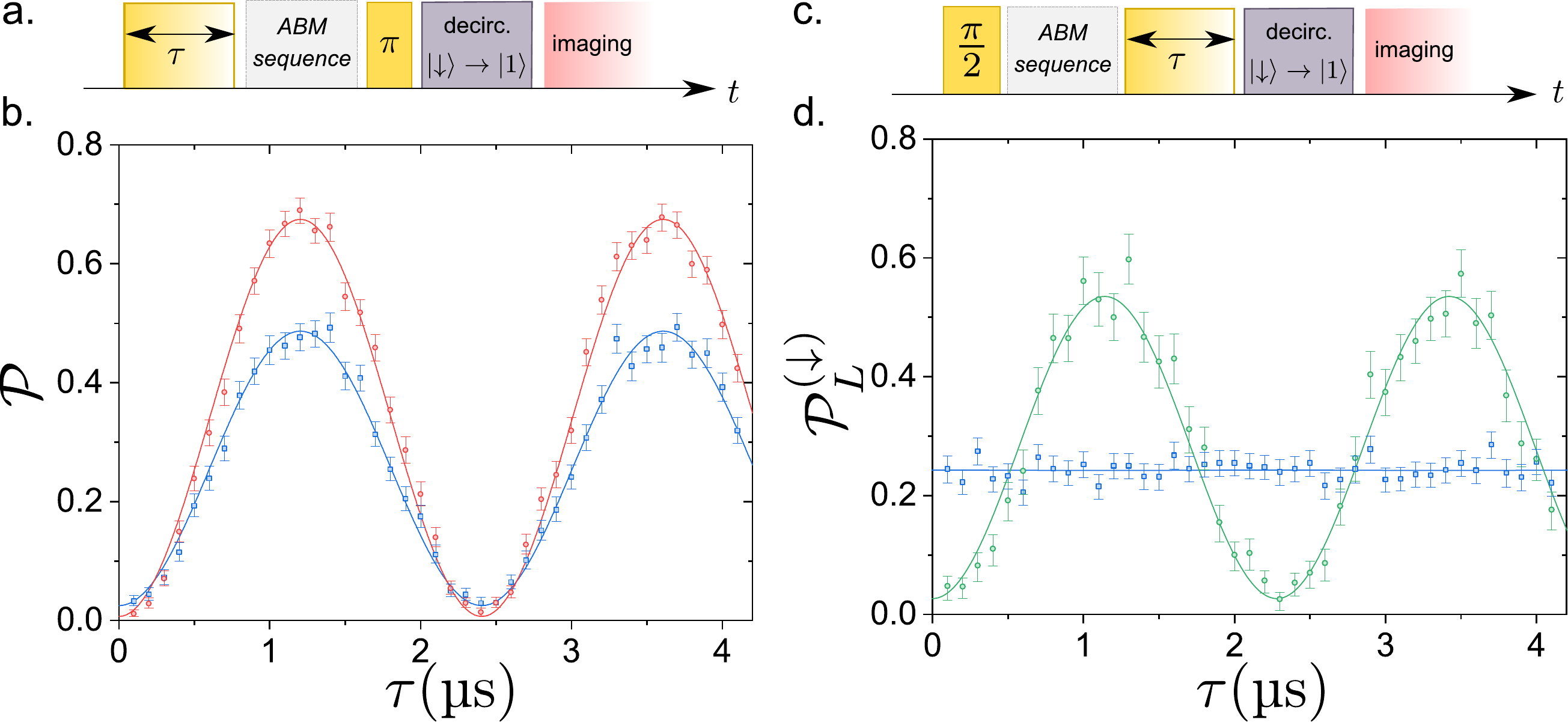}
\caption{\textbf{Assessing the QND character of the ABM with Rabi oscillations.} \textbf{a.} Pulse sequence used to record Rabi oscillations on the $\kdown\to\kup$ transition. After a MW pulse of duration $\tau$, the population in $\kup$ is measured with the ABM and the DM schemes. The color code is that of Fig.~\ref{fig:Spectrum}.a. \textbf{b.} Probabilities $\prob_{AA}$ and $\probc^{(\Sup)}$ to recapture the ancillae (red circles) and the logical atom (blue squares), respectively, measured with the sequence sketched in panel \textbf{a}. The solid lines are sinusoidal fits to the data with a shared oscillation frequency and fitted contrasts of $\num{0.668\pm0.007}$ and $\num{0.461\pm0.004}$ for the ABM and DM, respectively. \textbf{c.} Pulse sequence used to record Rabi oscillations with the DM, set to measure $\kdown$. The oscillations are driven after a $\pi/2$ pulse that prepares the logical atom in $(\kup+\kdown)/\sqrt{2}$ and an ABM of the logical state. The color code is that of Fig.~\ref{fig:Spectrum}.a. \textbf{d.} Probability $\probc^{(\Sdown)}$ to recapture the logical atom averaged over all realizations of the experiment (blue squares), i.e, unconditioned on the ABM outcome, or conditioned on the recapture of the ancilla, i.e., on the non-destructive measurement of $\kup$ (green circles). The solid lines are sinusoidal fits to the data with shared oscillation frequencies and fitted contrasts of $\num{0.51\pm0.01}$ and $\num{0.5\pm5}\times10^{-4}$. In panels \textbf{b} and \textbf{d}, the data points are the result of an average over 200 realizations of the experiment, the error bars are statistical.}
\label{fig:Rabi}
\label{fig:QND}
\end{figure*}   

We use the ABM sequence with a single ancilla to measure the state of the logical atom, after its preparation in $\kdown$ or $\kup$. The effective preparation of these states is checked by a DM of $\ket{\psi_L}$, which we perform after step (iv) of the ABM scheme. The final fluorescence image, which equally detects atoms in $\kgo$ or $\kgt$, reads the outcome of both the DM and ABM. %

In Figure~\ref{fig:Spectrum}.b, we plot the probabilities to detect the ancilla conditioned on the destructive detection of the logical qubit in $\kdown$ and $\kup$, denoted $\prob_{A|\Sdown}$ and $\prob_{A|\Sup}$ respectively, as a function of the detuning $\Delta$ of the ancilla optical excitation from the $\kgt\to\kR$ transition. %
The probability $\prob_{A|\Sdown}(\Delta)$ (blue squares) displays a coherent excitation profile, which reveals the excitation of the ancilla to $\kR$ when $\Delta=0$, unaltered by the presence of the logical atom in $\kdown$. 
The symmetric splitting of $\prob_{A|\Sup}(\Delta)$ (red circles) around $\Delta=0$ is a clear signature of the interaction with the logical atom in $\kup$: the excitation to $\kR$ at resonance is essentially suppressed. The residual excitation at $\Delta=0$ is the signature of a false-positive DM of the logical atom in $\kup$, which may stem from a decay to $\kgo$ of the low-$\ell$ Rydberg levels transiently populated during the circular state preparation~\cite{Ravon2023}. The two main peaks correspond to the excitation to $\ket{-}$ or $\ket{+}$. From their splitting $\delta\nu = 2V/h = \SI{19.6\pm0.1}{\mega\hertz}$, we find an interatomic distance $d=\SI{6.12\pm0.01}{\micro\meter}$ close to the $\SI{6}{\micro\meter}$ programmed value~\cite{Mehaignerie2025}.  %
Their widths ($\SI{4}{\mega\hertz}$) is found to be mainly due to the thermal spread of the atomic positions within the traps~\cite{Suppl}. 
At $\Delta=0$, the mapping by the ABM sequence of the circular states $\kdown$ and $\kup$ onto the ancilla states $\kgo$ and $\kgt$ is conspicuous: the ancilla is detected in $\kgo$ only if the logical atom has been prepared in $\kup$. From the data of Figure~\ref{fig:Spectrum}.b, we estimate the false negative probability $1-\prob_{A|\Sup}(\Delta=0)$ to be about $\SI{6.7}{\percent}$ and false positive probability $\prob_{A|\Sdown}(\Delta=0)\approx\SI{1.3}{\percent}$.  

In the case where we prepare the logical atom in $\kup$, we plot in Fig.~\ref{fig:Spectrum}.c the probability $\probc^{(\Sup)}$ to recapture the logical atom after the final DM of $\kup$. The probabilities are the same whether $\Delta=0$ (full ABM) or $|\Delta|>\SI{15}{\mega\hertz}$ (no ABM), showing that the state of the logical atom is unaltered by the ABM. The reduction of $\probc^{(\Sup)}$ when $\Delta\sim\pm V/h$ corresponds to the excitation of the $\ket{\pm}$ states. A fraction of the logical state then overlaps with $\kaux$ and is not detected by the DM.

We verify that the ABM fulfils the requirements of a genuine QND measurement by using it to record Rabi oscillations of the logical atom. We initially prepare an ensemble with two ancillae in the $\ket{\Sgt, \Sdown, \Sgt}\ALA$ state. We apply a MW pulse of variable duration, $\tau$ (see Fig.~\ref{fig:Rabi}.a), resonant with the $\kdown\to\kup$ transition and measure the state $\kup$ with the DM and the ABM schemes. 
We plot in Fig.~\ref{fig:Rabi}.b, as a function of $\tau$, in blue $\probc^{(\Sup)}$ and in red the probability to recapture both ancillae, denoted $\prob_{AA}$. Rabi oscillations are apparent in both signals, showing that the ABM reproduces the evolution of $\ket{\psi_L}$ found by the DM. Remarkably, the contrast of the ABM scheme is notably higher than that of the DM scheme. The ABM is not affected by the $\SI{66}{\percent}$ efficiency of the transfer from $\kdown$ to $\kgt$, which corresponds to the detection efficiency of the DM~\cite{Suppl}. 
The ABM contrast of $\SI{67.3\pm.8}{\percent}$ is essentially limited by the $\SI{72.3\pm0.4}{\percent}$ preparation efficiency of $\kdown$ rather than by the ABM detection efficiency. %

\begin{figure*}[t]
\centering
\includegraphics[width=1.9\columnwidth]{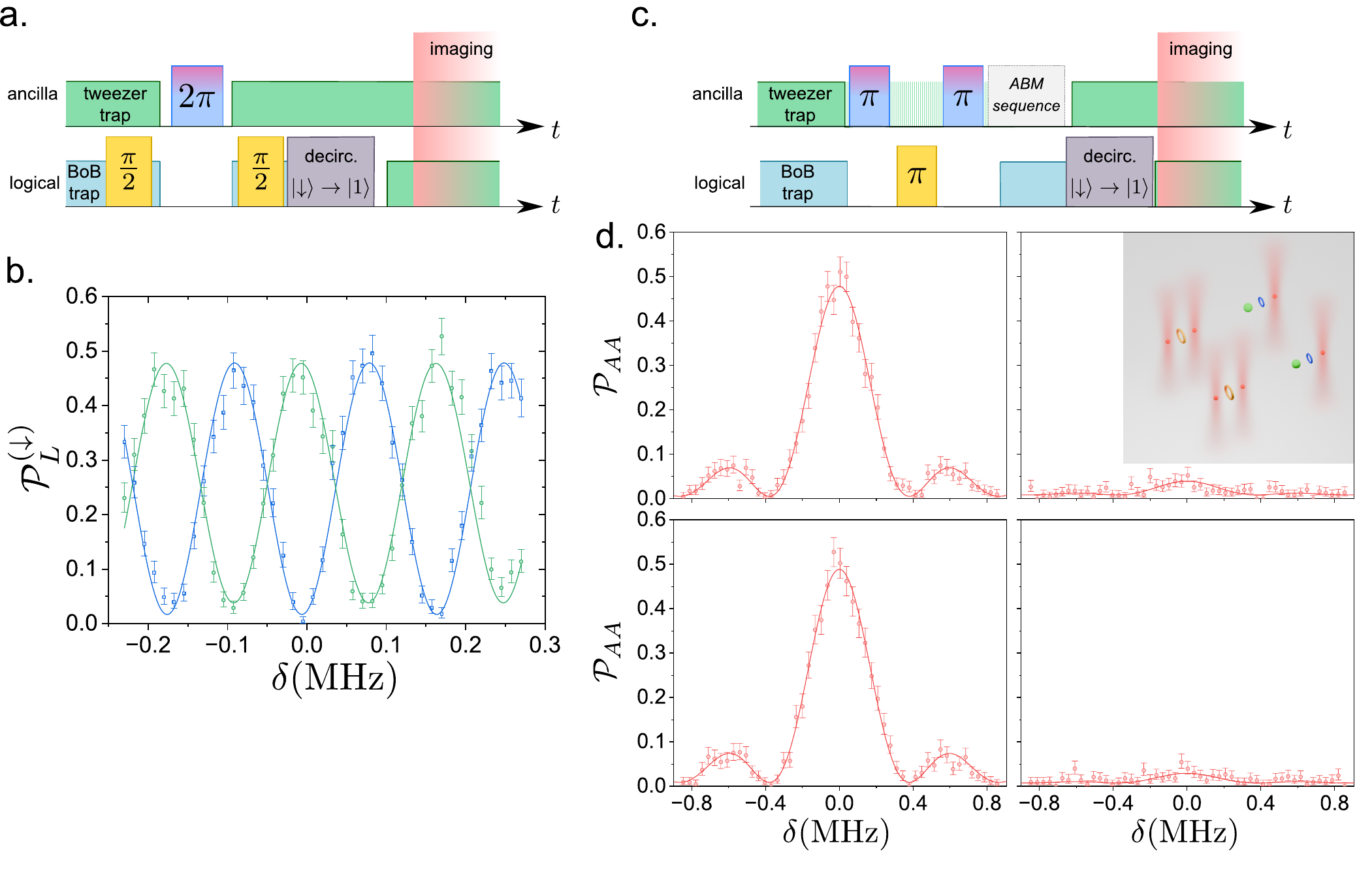}
\caption{\textbf{Local manipulation of the logical atom with the ancilla.} \textbf{a.} Pulse sequence used to measure a $\pi$ phase shift on the logical atom using a $2\pi$ optical pulse on the ancilla. The two MW $\pi/2$ pulses define a Ramsey interferometer. The colour code is that of Fig.~\ref{fig:Spectrum}.a. \textbf{b.} Probability to measure the logical atom in $\kdown$ with the destructive scheme after the Ramsey sequence of panel \textbf{a}, as a function of the frequency of the MW field. The optical $2\pi$ pulse on the ancilla atom is either applied (green circles) or absent (blue squares). The points are the result of an average over 100 repetitions, the error bars are statistical. The solid lines are fits of a sine law to the data with contrasts $\num{0.46\pm0.01}$ (green) and $\num{0.44\pm0.01}$ (blue) and a $\num{1.01\pm0.02}\pi$ phase difference. \textbf{c.} Pulse sequence used to locally manipulate the state of the logical atom. We apply a MW $\pi$ pulse on the logical atom in between two optical $\pi$ pulses on the ancillae, during which the tweezer traps are turned off (striped green area) for the control ancillae. \textbf{d.} (Inset) Structure of the array of four ensembles of logical and ancilla atoms. The two ancillae, the tweezer traps of which are turned off during the optical $\pi$ pulses of panel \textbf{c}, are excited to $\kR$ (green sphere) while the others stay in $\kgt$ (red sphere). The two corresponding logical atoms on the right stay in $\kdown$ (blue torus) while the two left ones are transferred to $\kup$ (yellow torus). Probability (red circles) to measure the logical atom in $\kdown$ with the ABM, plotted as a function of the frequency of the MW field. Each panel corresponds to the ensemble with the corresponding position in the inset. The points are the result of an average over 400 repetitions, the error bars are statistical. The solid red lines are fits of a spectral line profile to the data~\cite{Suppl}, with amplitudes, from left to right and top to bottom, of $\num{.48\pm.01}$, $\num{.03\pm.01}$, $\num{.485\pm.07}$ and $\num{.022\pm.005}$.}
\label{fig:manipulation}
\end{figure*}

To assess the ABM fidelities without contamination of the DM level crosstalks which affect previous estimations from Fig.~\ref{fig:Spectrum}~\cite{Suppl}, we measure the joint probabilities to detect none, part of or all of the three atoms in the fluorescence imaging that follows the Rabi oscillations, for two sequences in which we perform a DM of $\kup$ (as in Fig.~\ref{fig:Rabi}.b) or $\kdown$. We define $\eta_\Sup$ ($\eta_\Sdown$) as the probability for the ancilla to stay in $\kgt$ (to be transferred to $\kR$), at step (ii) of the ABM sequence, when the logical qubit is in $\kup$ ($\kdown$). 
From an experimental-imperfection model fitted to the measured Rabi oscillations~\cite{Suppl}, we get $\eta_\Sup = \SI{96.6\pm0.2}{\percent}$ and $\eta_\Sdown = \SI{95.0\pm0.3}{\percent}$. The efficiency $\eta_\Sup$ is limited by ancilla losses due to collisions with the background gas or to finite recapture efficiency. The main limitations to $\eta_\Sdown$ are the finite excitation efficiency of $\kR$ for an isolated atom and the lifetime of $\kR$. These technical limitations may be reduced in an improved setup with, e.g., a cryogenic environment to increase the lifetimes of both the atoms in their traps and of the Rydberg levels. 

Finally, we check that the ABM possesses the projective character of a QND measurement. After a MW $\pi/2$ pulse that prepares the logical atom in $(\kdown+\kup)/\sqrt{2}$, we perform the ABM sequence with a single ancilla, apply a second variable-duration MW pulse that drives Rabi oscillations (see Fig.~\ref{fig:QND}.c) and finish with the DM of the logical atom. 
Discarding the ABM information, we get $\probc^{(\Sdown)}(\tau)$ and plot it in Fig.~\ref{fig:QND}.d. In spite of the second MW pulse, we find no oscillations, revealing the maximal entanglement between the logical atom and the ancilla. However, if we select only the experimental realizations where the ABM has measured the logical atom in $\kup$, we recover fully contrasted oscillations of $\probc^{(\Sdown)}(\tau)$, starting from $\probc^{(\Sdown)}(0)={\num{0.01\pm0.01}}$: The ABM has projected the logical atom onto $\kup$,  as expected for a genuine QND measurement. Remarkably, this also shows that the ABM can be used to measure $\ket{\psi_L}$ at any time in the sequence and thus to operate mid-circuit measurements. Note that the effective duration of the QND measurement corresponds to the $\SI{2.5}{\micro\second}$ duration of steps (ii) to (iv). It is not limited by the $\SI{25}{\milli\second}$-long fluorescence imaging at the end of the experimental sequence.

We now use the ancilla to optically apply phase and spin flips to the logical atom. An optical $2\pi$ pulse on the $\kgt\to\kR$ ancilla transition maps the states $\ket{\Sdown, \Sgt}\LA$ and $\ket{\Sup, \Sgt}\LA$ to $-\ket{\Sdown, \Sgt}\LA$ and $\ket{\Sup, \Sgt}\LA$, respectively. When $\ket{\psi_L}=\kup$, the Rydberg blockade of the ancilla excitation (Fig.~\ref{fig:setup}.b) inhibits the $\pi$ phase shift experienced by $\ket{\Sdown, \Sgt}\LA$, which we probe with a MW Ramsey interferometer (see Fig.~\ref{fig:manipulation}.a). In Fig.~\ref{fig:manipulation}.b, we plot $\probc^{(\Sdown)}$ measured after two MW $\pi/2$ pulses on the $\kdown\to\kup$ transition separated by $\SI{3}{\micro\second}$, as a function of the detuning, $\delta$, of the driving MW field from resonance. The Ramsey fringes are measured either with (green) or without (blue) the optical $2\pi$ pulse on the ancilla. The expected $\pi$ phase shift is conspicuous. A sine fit to the data reveals a $\SI{1.01\pm0.02}{\pi}$ phase shift with a contrast reduction by a factor of only $\num{0.93\pm0.04}$. This demonstrates our ability to optically imprint a $\pi$ phase shift on the circular-state-encoded logical qubit, naturally impervious to optical fields. %

We also demonstrate an optically-controlled spatially-resolved spin flip of the logical atom by using an ancilla excited to $\kR$, which blocks the MW transfer of the logical atom between $\kdown$ and $\kup$ (Fig.~\ref{fig:setup}.b). 
We prepare an array containing four logical atoms, initially excited to $\kdown$, each coupled to two ancillae, as drawn in the inset of Fig.~\ref{fig:manipulation}.d. One of the ancillae is used for the control of the logical atom state, and both of them for the measurement of this state. We block the transition from $\kdown$ to $\kup$, induced by a MW $\pi$ pulse  (see Fig.~\ref{fig:manipulation}.c for the pulse sequence), in the two ensembles on the right by exciting the control ancillae to $\kR$ before the MW pulse. 
To do so, we turn off the optical tweezers during the excitation to $\kR$ for these two ancillae only~\cite{Suppl}. The six other ancillae remain in $\kgt$ because of the trap-induced light shifts. After the MW pulse, a second optical $\pi$ pulse brings the two control ancillae back to $\kgt$. 
We eventually turn back on all the tweezer traps. This leaves the two right (left) logical atoms in $\kdown$ ($\kup$) and all ancillae in $\kgt$. Remarkably, the ancillae, including the control ones, can be used again to perform a QND measurement of the state of the logical atom. 
 We plot in Fig.~\ref{fig:manipulation}.d $\prob_{AA}$ for the four ensembles of the array, as a function of the MW frequency detuning $\delta$, exhibiting the excitation spectrum of the $\kdown\to\kup$ transition. While the left atoms end up in $\kup$ when $\delta=0$, the MW excitation of the right atoms is strongly suppressed. The residual excitation results from the finite efficiency of the initial excitation of the ancilla to $\kR$~\cite{Suppl}.

We have developed a hybrid dual-Rydberg platform that enables high-fidelity quantum non-demolition measurements and spatially-resolved optical manipulations of individual circular Rydberg atoms. The measurement time is in the microsecond range, well matched to the timescales of the circular-Rydberg-atom platforms~\cite{Nguyen2018, Cohen2021}. It could be reduced by replacing steps (ii-v) of the ABM sequence with a $2\pi$ pulse on the $\kgt\to\kR$ transition, which effectively operates a logical-atom-controlled-Z gate on the ancilla, within a Ramsey interferometer on the ground-state hyperfine transition. The interferometer would read the phase shift and, thus, the state of the logical atom. This would in addition reduce the imperfections due to the finite lifetime of the low-$\ell$ $\kR$ state. 

In our experiment, we use up to two ancillae per logical atom. Using an assembly of ancilla atoms could increase the measurement fidelity~\cite{Petrosyan2024}. It would also enable the detection of multiple circular Rydberg levels or the measurement of the same atom at different times. The ancillae could also be used to mediate rapidly tunable interactions between logical atoms~\cite{Botzung2025}, even at a distance hence increasing the system connectivity~\cite{Nguyen2023}. 
Finally, our method is not restricted to the two circular Rydberg states employed in this work~\cite{Suppl}. It can be readily extended to quantum simulators that involve other non-circular high-angular-momentum Rydberg levels~\cite{Kruckenhauser2022}, using the wide Stark tunability of Rydberg levels to reach a Förster resonance, or circular Rydberg states of other species than alkali atoms, e.g., strontium~\cite{Teixeira2020, Holzl2024}. 

This hybrid dual-Rydberg platform is very promising for quantum simulation. It allows for the initialization of the simulator in an arbitrary product state. The QND nature of the detection allows one to measure correlations between arbitrary single-qubit observables at different times, including out-of-time-order correlations~\cite{Swingle2018, Lewis-Swan2019, Xu2024, Liang2025}. This is a crucial feature to take full benefit of the long simulation times enabled by circular Rydberg atoms~\cite{Nguyen2018}. It opens the route to the study of long-term coherent evolutions, slow thermalization dynamics~\cite{Turner2018a, Serbyn2021} and to the observation of measurement-induced phase transitions~\cite{Gullans2020,Turkeshi2021,Fresco2024}. %

\medskip
This publication has received funding by the France 2030 programs of the French National Research Agency (Grant No. ANR-22-PETQ-0004, project QuBitAF), under Horizon Europe programme HORIZON-CL4-2022-QUANTUM-02-SGA via the project 101113690 (PASQuanS2.1), by the European Union (ERC Advanced Grant No. 786919, project TRENSCRYBE). It has been supported by the Île-de-France region in the framework of DIM QuanTiP (project LT-CRAQS) and by the Quantum Information Center Sorbonne as part of the program \emph{Investissements d'excellence} -- IDEX of the Alliance Sorbonne Université.

\bibliography{Machu2025}

\end{document}